\documentclass[
    ,final            
  ]
  {aipproc}
\usepackage{graphicx}

\layoutstyle{6x9}

\def \Kla#1{\left( #1 \right)}
\def \SCIe#1#2{\hbox{$#1\times 10^{#2}$}}

\begin{document}

\title{Neutron Beta Decay Studies with Nab}


\newcommand{\UVA}{Physics Department, University
             of Virginia, Charlottesville, VA 22904, USA} 
\newcommand{\ORNL}{Physics Division, Oak Ridge National Laboratory, Oak Ridge, TN 37831, USA} 
\newcommand{\ASU}{Department of Physics, Arizona State University, Tempe, AZ 85287-1504, USA}
\newcommand{\Sussex}{Department of Physics and Astronomy, University of Sussex, Brighton BN19RH, UK}
\newcommand{\UNH}{Department of Physics, University of New Hampshire, Durham, NH 03824, USA}
\newcommand{\UKy}{Department of Physics and Astronomy, University of Kentucky, Lexington, KY 40506, USA}
\newcommand{\Manitoba}{Department of Physics, University of Manitoba, Winnipeg, Manitoba, R3T 2N2, Canada}
\newcommand{\UMich}{University of Michigan, Ann Arbor, MI 48109, USA}
\newcommand{\KIT}{IEKP, Universität Karlsruhe (TH), Kaiserstra\ss{}e 12, 76131 Karlsruhe, Germany}
\newcommand{\UT}{Department of Physics and Astronomy, University of Tennessee, Knoxville, TN 37996, USA}
\newcommand{\USC}{Department of Physics and Astronomy, University of South Carolina, Columbia, SC 29208, USA}
\newcommand{\LANL}{Los Alamos National Laboratory, Los Alamos, NM 87545, USA}
\newcommand{\Winnipeg}{Department of Physics, University of Winnipeg, Winnipeg, Manitoba R3B2E9, Canada}
\newcommand{\UNC}{Department of Physics, North Carolina State University, Raleigh, NC 27695-8202, USA}
\newcommand{\Mexico}{Universidad Nacional Aut\'onoma de México, México, D.F. 04510, M\'exico}
\newcommand{\NCSU}{Physics Department, NC State University, Raleigh, NC 27695, USA}

\author{S.~Bae\ss{}ler}{address=\UVA,altaddress=\ORNL}
\author{R.~Alarcon}{address=\ASU}
\author{L.P.~Alonzi}{address=\UVA}
\author{S.~Balascuta}{address=\ASU}
\author{L.~Barr\'on-Palos}{address=\Mexico}
\author{J.D.~Bowman}{address=\ORNL}
\author{M.A.~Bychkov}{address=\UVA}
\author{J.~Byrne}{address=\Sussex}
\author{J.R.~Calarco}{address=\UNH}
\author{T.~Chupp}{address=\UMich}
\author{T.V.~Cianciolo}{address=\ORNL}
\author{C.~Crawford}{address=\UKy}
\author{E.~Frle\v{z}}{address=\UVA}
\author{M.T.~Gericke}{address=\Manitoba}
\author{F.~Gl\"uck}{address=\KIT}
\author{G.L.~Greene}{address=\UT,altaddress=\ORNL}
\author{R.K.~Grzywacz}{address=\UT}
\author{V.~Gudkov}{address=\USC}
\author{D.~Harrison}{address=\Manitoba}
\author{F.W.~Hersman}{address=\UNH}
\author{T.~Ito}{address=\LANL}
\author{M.~Makela}{address=\LANL}
\author{J.~Martin}{address=\Winnipeg}
\author{P.L.~McGaughey}{address=\LANL}  	
\author{S.~McGovern}{address=\UVA}
\author{S.~Page}{address=\Manitoba}
\author{S.I.~Penttil\"a}{address=\ORNL}
\author{D.~Po\v{c}ani\'c}{address=\UVA}
\author{K.P.~Rykaczewski}{address=\ORNL}
\author{A.~Salas-Bacci}{address=\UVA}
\author{Z.~Tompkins}{address=\UVA}
\author{D.~Wagner}{address=\UKy}
\author{W.S.~Wilburn}{address=\LANL}
\author{A.R.~Young}{address=\NCSU}


\begin{abstract}
Precision measurements in neutron beta decay serve to determine the coupling constants of beta decay and allow for several stringent tests of the standard model. This paper discusses the design and the expected performance of the Nab spectrometer.


\end{abstract}

\maketitle



A program of precision studies of neutron beta decay is planned with the Nab spectrometer. Nab will use the state of the art neutron beamline (FNPB) at the new Spallation Neutron Source (SNS) in Oak Ridge, TN. The expected results will play a critical role in resolving longstanding discrepancies in the neutron decay world data set, and will allow an extraction of $V_{\rm ud}$, the upper left element of the Cabbibo-Kobayashi-Maskawa matrix.
The redundancy inherent in the standard model description of the neutron beta decay process allows uniquely sensitive checks of the model's validity and limits \cite{Abe08,Nico09,Kon10,Bhat12}, with strong implications in astrophysics \cite{Dub11}.

The aim of the Nab collaboration is to determine $a$, the neutrino electron correlation coefficient, and $b$, the Fierz parameter, 
in the decay of the free neutron \cite{Jack57}.
The $a$ coefficient can be obtained through a measurement of the electron energy $E_{\rm e}$, and the momentum $p_{\rm p}$ of the corresponding proton. 
In zeroth order approximation, the $a$ coefficient is given by the slope of the top of the squared proton momentum distribution $P_{\rm p}(p_{\rm p}^2)$ for a fixed electron energy $E_{\rm e}$:
\begin{equation}
P_{\rm p}(p_{\rm p}^2) \propto \left\{
\begin{array}{cl}
\Kla{1+a\frac{p_{\rm e}}{E_{\rm e}}\frac{p_{\rm p}^2-p_{\rm e}^2-p_\nu^2}{2 p_{\rm e} p_\nu}}
& \textrm{if } -1 \le \frac{p_{\rm p}^2-p_{\rm e}^2-p_\nu^2}{2 p_{\rm e} p_\nu} \le 1\\
0 & \textrm{otherwise } \\
\end{array} \right.
\end{equation}
In this approximation, electron ($p_{\rm e}$) and neutrino ($p_\nu$) momenta are functions of the electron energy $E_{\rm e}$. The positions of the edges of these distributions $P_{\rm p}(p_{\rm p}^2)$ are sharply defined by the three-body decay kinematics.
The Fierz term $b$ could in principle be determined from a precise measurement of the beta energy spectrum alone. However, the Nab collaboration believes that coincident detection of the accompanying proton is essential in suppressing background.
The basic principles of the Nab spectrometer have been described in Refs. \cite{Bow05,Ala07, Poc09}. This paper reports on a conceptual design update. 



\begin{figure}[!ht]%
  \parbox[t]{0.45\textwidth}{%
  {\centering \includegraphics{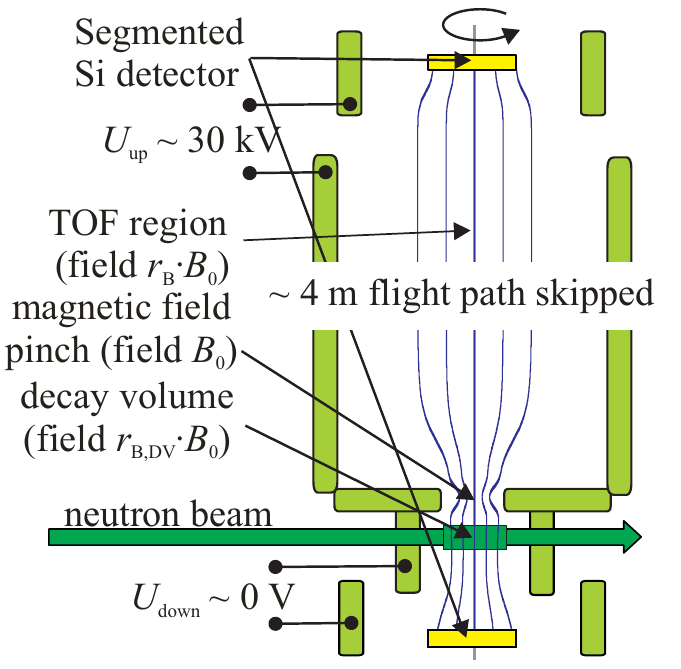}}
   }%
  \qquad
  \begin{minipage}[t]{0.45\textwidth}
    {\centering \includegraphics{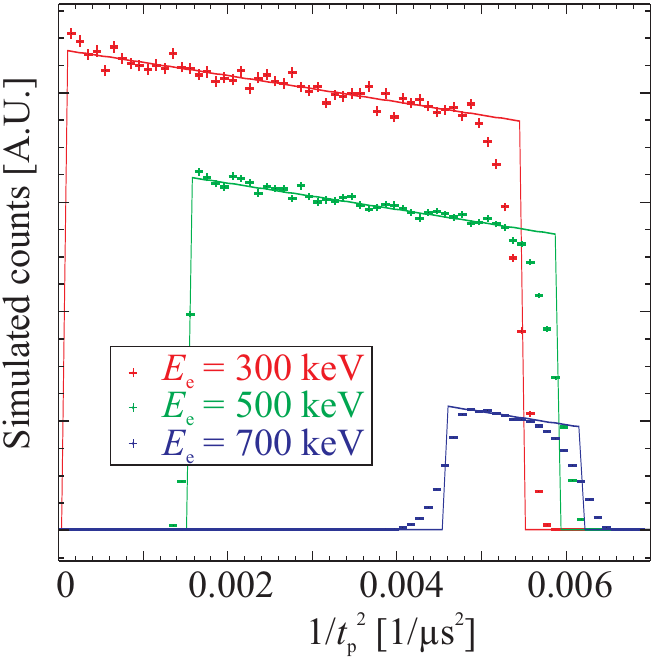}} 
    \caption{Left: Sketch of the Nab spectrometer setup, not to scale. Magnetic field lines (shown in blue) electrodes (light green boxes), and coils (not shown) possess cylindrical symmetry around the vertical axis. Right: Simulated inverse squared proton $1/t_{\rm p}^2$ histograms for several given electron energies $E_{\rm e}$ are given as colored dots. The straight lines show an ideal detection system with $p_{\rm p}^2 \propto 1/t_{\rm p}^2$.} 
    \label{fig:All}
  \end{minipage}%
\end{figure}%

A sketch of the Nab spectrometer is shown in Fig.~\ref{fig:All}. The Si detectors measure the electron energy with keV-level resolution. Electron energy losses through backscattering are avoided thanks to the magnetic guide field that connects the two detectors at both sides of the decay volume. Electrons might bounce, but are eventually fully absorbed. Their full energy will be determined after adding signals from both detectors. Corrections for dead-layer(s) and bremsstrahlung will be made.
The proton is detected only if it arrives at the upper detector, after a long TOF path and subsequent acceleration. Its momentum $p_{\rm p}$ is inferred from the proton's time of flight $t_{\rm p}$ relative to the electron hit. Once the proton momentum is longitudinalized by the magnetic field, $p_{\rm p} \propto 1/t_{\rm p}$, but a certain path length is required for that.

Also shown in Fig.~\ref{fig:All} is the simulated performance of a realistic electromagnetic design of the spectrometer, and compared to an ideal detector response with $p_{\rm p}^2 \propto 1/t_{\rm p}^2$. Electric field effects are neglected.
The edges of the simulated $1/t_{\rm p}^2$ distribution reflect the detector response function. In comparison to the previous, symmetric design (see Fig.~3 in \cite{Poc09}), the relative width of the response function is considerably smaller in the asymmetric configuration, while maintaining a workable count rate.
The bigger flight path length and the sharper magnetic field pinch outweigh the fact that the protons have to pass through the full field pinch. 

The Si detector was initially designed by the abBA collaboration \cite{Ala07b}, and is being refined by the Nab and UCNB collaborations. For first performance studies, see \cite{Sal12}.

\begin{table}[!ht]
\begin{tabular}{lc}
     \hline
     \textbf{Experimental parameter} & \textbf{Systematic uncertainty $\Delta a/a$} \\
     \hline
      Magnetic field curvature at pinch & \SCIe{5}{-4}\\
      $\dots$ ratio $r_{\rm B}=B_{\rm TOF}/B_0$ & \SCIe{2.5}{-4} \\
      $\dots$ ratio $r_{\rm B,DV}=B_{\rm DV}/B_0$ &  \SCIe{3}{-4} \\
      Electrical potential inhomogeneity in decay volume / filter region & \SCIe{5}{-4} \\
      $\dots$ in TOF region & \SCIe{1}{-4} \\
      Neutron Beam position & \SCIe{4}{-4} \\
      $\dots$ profile (including edge effect) & \SCIe{2.5}{-4} \\
      Adiabaticity of proton motion & \SCIe{1}{-4}\\
      Detector: Electron energy resolution & \SCIe{5}{-4}  \\
      $\dots$ Proton trigger efficiency & \SCIe{2.5}{-4} \\
      \hline
      Sum & \textbf{\SCIe{1}{-3}} \\
\hline
\end{tabular}
  \caption{Predicted uncertainties in $a$
 due to imperfect knowledge of spectrometer properties.
 }
  \label{tab:SystUncertainty_a}
\end{table}

Counting statistics will not be limiting: After about six weeks of data taking, the statistical uncertainty in the determination of $a$ is below $0.1\%$. Tab.~\ref{tab:SystUncertainty_a} shows the predicted systematic uncertainties. Several additional items (length of TOF region, electron energy calibration, residual gas, background, accidental coincidences, unwanted neutron beam polarization and Doppler effect) have been found to be small if the relevant specifications are met.
See \cite{Ala10} for a detailed discussion.

We gratefully acknowledge funding support from NSF and DOE.





\bibliographystyle{aipproc}   

\end{document}